\documentclass[12pt,preprint]{aastex}

\shorttitle{Inner-Halo RRL Overdensity}

\begin{document}

\title{A Kinematically-Distinct RR-Lyrae Overdensity in the Inner Regions of the Milky Way}

\author{Dana I. Casetti-Dinescu\altaffilmark{1,2}, 
Daniel A. Nusdeo\altaffilmark{1}, 
Terrence M. Girard\altaffilmark{2}, 
Vladimir I. Korchagin\altaffilmark{3} and 
William F. van Altena\altaffilmark{2}}

\altaffiltext{1}{Department of Physics, Southern Connecticut State University, 501 Crescent St., New Haven, CT 06515, USA, casettid1@southernct.edu,~nusdeod1@owls.southernct.edu}
\altaffiltext{2}{Astronomy Department, Yale University, 260 Whitney Ave. , New Haven, CT 06511, USA, 
dana.casetti@yale.edu,~terry.girard@yale.edu,~william.vanaltena@yale.edu}
\altaffiltext{3}{Institute of Physics, Southern Federal University, Stachki st. 194, 344090, Rostov-on-Don, Russia, vkorchagin@sfedu.ru}

\begin{abstract}
We combine the Siding Spring Survey of RR Lyrae stars with the Southern Proper Motion Catalog 4, in order
to detect and kinematically characterize overdensities in the inner halo of the Milky Way. 
We identify one such overdensity 
above the Galactic plane, in quadrant 4 of the Galaxy. The overdensity extends 
at least $20^{\circ}$ in longitude, has an average heliocentric 
distance of 8 kpc with a depth of 4 kpc, and is confined within 4 kpc of the Galactic plane.
Its metallicity distribution is distinct from that of the field population having
a peak at -1.3 and a pronounced tail to -2.0.
Proper motions indicate a net vertical motion away from the plane, and
a low orbital angular momentum. Qualitatively, these orbit properties suggest a 
possible association with $\omega$ Centauri's parent satellite. 
However, comparison to a specific $\omega$ 
Cen N-body disruption model does not give a good match with observations.
Line-of-sight velocities, and more extensive N-body modelling will help
clarify the nature of this overdensity.

\end{abstract}


\section{Introduction}

RR Lyrae stars are established tracers of old, metal-poor populations. Being bright,
well-calibrated standard candles, these stars have been used extensively 
for studies of the halo, thick disk and bar of our Galaxy, as well as studies of the Milky Way (MW) satellites. 
Typically, the presence (or absence) of
an RR-Lyrae stellar overdensity will help
establish (or refute) the origin of the overdensity as a disrupted satellite (see e.g.,
 Cseresjnes {\it et al.} 2000 for the Sagittarius dwarf galaxy,
Vivas \& Zinn 2006 for the Virgo Overdensity,
Mateu {\it et al.} 2009 for the Canis Major overdensity). 

The recent releases of large-area RR Lyrae databases by the Catalina surveys 
(Drake {\it et al.} 2013ab, Torrealba {\it et al}. 2015 - hereafter T15)
have opened new avenues of halo-substructure investigation. Here, we report on one such study in which
we combine the recently released 
Siding Spring Survey (SSS; aka Catalina south) of RR Lyrae stars described in T15
with the Southern Proper Motion Catalog SPM4 (fourth installment, Girard et al. 2011, hereafter G11). 
Our aim is to detect and kinematically characterize stellar overdensities 
within a heliocentric distance of 15 kpc, specifically in the inner halo/thick disk of the MW.

Comparing two areas symmetrically situated about the Galactic plane, we find an overdensity in the fourth 
quadrant of the Galaxy, above the plane. This overdensity was also singled out by T15 in their comparison with a halo
density model; however it was unclear whether that detection was the result of a poor halo model, or a genuine
stellar excess. Here we establish that the overdensity is kinematically distinct from canonical MW components.
Specifically we find that the overdensity:
1) has a net vertical motion, yet it is confined to within 4 kpc of the plane, 2)
has a relatively low orbital angular momentum, allowing for both prograde and retrograde orbits and 3)
is predominantly metal-poor with a peak at [Fe/H] = -1.3 and a pronounced tail to [Fe/H] = -2.0.

Qualitatively, our findings are consistent with debris from $\omega$ Centauri's parent satellite: i.e. a
flattened spatial distribution with low orbital angular momentum (Tsuchiya et al. 2003, 2004; hereafter T04, 
Bekki \& Freeman 2003, Mizutani et al. 2003, Meza et al. 2005). However,
a direct comparison with the T04 N-body model does not give a good match. 
Likewise, a comparison with 
the Law \& Majewski (2010) model for the Sagittarius dwarf galaxy does not favor such an origin.
We speculate that another possible origin mechanism is bar-induced resonant trapping as envisioned by Moreno et al. (2015). 

Key to clarifying the origin of this overdensity are radial velocities: N-body models 
for $\omega$ Cen debris give an average velocity that is distinct from field stars. Also, better 
constrained models should help clarify the origin, in particular the above- versus below-the-plane
asymmetry. 

\section{Catalog Data}

The SSS RR Lyrae catalog (T15) contains $\sim 10,000$ stars down to a limiting magnitude 
$V = 19$. 
The SSS detects ab-type RR Lyrae (RRab) within
the declination range $-75^{\circ} \le \delta \le -15^{\circ}$, and has 
an estimated overall completeness of $\sim 70\%$.
The SPM4 catalog covers the sky below $\delta \le -20^{\circ}$, and
 is approximately complete to $V \sim 17.5$ (G11). Thus, there is good overlap between the two 
catalogs. 

During the cross-matching process we found that the SSS positions 
have errors of up to 5 arcsec. Specifically, the separation in positions between 
SPM4 and SSS has a wide distribution that varies with declination; 1-2 arcsec 
at $\delta = -60^{\circ}$, increasing to 4-5 arcsec at $\delta = -20^{\circ}$. 
We therefore chose a matching tolerance of 5 arcsec, then eliminated duplicates/multiples by 
keeping the lowest separation. Objects with magnitude differences
in excess of 2.0 mags were discarded. This yielded a list of 8023 unique matches.
In Figure 1 we show the spatial distribution of these objects.

\section{Analysis}

\subsection{Sample Selection}

We focus on two regions symmetrically placed with respect to the Galactic plane and with
longitude range $-40^{\circ} \le l \le 0^{\circ}$. This choice was prompted by the area-coverage uniformity 
in the above- and below-galactic plane samples; hereafter designated {\it ABOVE} and {\it BELOW}. 

The ``primary'' samples have a latitude range of $18^{\circ} \le |b| \le 30^{\circ}$ (Fig.1 continuous line), 
and these were used 
to produce spatial and metallicity distributions. The ``extended'' samples have a latitude range of
 $15^{\circ} \le |b| \le 30^{\circ}$ (Fig. 1, dashed line); these were used for the kinematical 
analysis.
At low latitudes, the SSS completeness varies with latitude due to crowding (see Fig. 1). Therefore,
we decided to
limit the spatial analysis to a low-latitude limit of $|b| = 18^{\circ}$ thus alleviating 
completeness variations in the {\it ABOVE} versus {\it BELOW} samples. 
For the kinematical analysis, we extend the low-latitude limit 
to $|b| = 15^{\circ}$.
Note that the {\it ABOVE} sample is missing a corner at the $l=0^{\circ}$ end, which can be mirrored
in the {\it BELOW} sample.

To minimize the effect of proper-motion errors, we restrict our analysis to objects within a heliocentric distance 
of $1 \le d \le 15$ kpc. With this distance limit, the sample is well-represented to $V=16.4$, with only
$0.4\%$ of the stars within $17 \le V \le 18$. 
Estimated proper-motion errors in SPM4 are $\sim 2-3$ mas/yr at these magnitudes
(see Fig. 6 in G11).

The stars we study are located in quadrant 4 of the Galaxy, to distances that 
well encompass the Galactic bulge. However, the low-latitude limit $|b| = 15^{\circ}$, corresponds to 
a vertical distance from the plane of 2.1 kpc at the center of the Galaxy ($R_0 = 8.0$ kpc). The plane-projected 
galactocentric radius $R_{GC}$ ranges between 0.9 and 9.0 kpc for the {\it ABOVE} sample,
and between 0.1 and 8.8 kpc for the {\it BELOW} sample. The near-end of the Galactic bar is located
in quadrant 1. Given our low-latitude limit of $15^{\circ}$ we therefore believe we do not sample
any portions of the known bar of the Milky Way (e.g., Wegg et al. 2015), and very little, if any of the bulge.

\subsection{Spatial and Metallicity Distributions}

With these cuts, we obtain 710 stars in the {\it ABOVE} sample, and 629 in the {\it BELOW} sample.
When the {\it BELOW} sample is trimmed to
exactly match the area in the {\it ABOVE} sample, i.e., subtract the corner at the $l=0^{\circ}$ end,
the number of stars is 473. 
The {\it ABOVE} sample has higher reddening ($E_{(B-V)} \le 0.4$) than that of
the {\it BELOW} sample ($E_{(B-V)} \le 0.2$) on average, and also a wider reddening distribution.
Expectations of star-counts are therefore opposite to the observations.

In Figure 2 we show the distributions in heliocentric distance, $d$; distance from the plane, $|Z|$; 
and metallicity, [Fe/H],
for the {\it ABOVE} (red) and the corner-trimmed {\it BELOW} (blue) samples. In the same plots, we also show the 
difference between the {\it ABOVE} and {\it BELOW} distributions by simple subtraction of one from another (black line). 
The frequency distributions are constructed using a box of half-width 1 kpc; 0.5 kpc; and 0.15 dex
in heliocentric distance; distance from the plane; and metallicity, respectively.
In all three plots, a stellar overdensity is apparent in the {\it ABOVE} sample. A normalization 
issue between the {\it ABOVE} and {\it BELOW} plane may arise simply because the {\it BELOW} sample may have 
fewer stars due to incompleteness differences at low latitudes. 
This is suggested by Fig. 2-top, where the difference shows a substantial
overdensity between distances of 6 to 10 kpc, while remaining flat yet
nonzero outside of this range.
The $|Z|$-distribution difference (middle panel) shows the overdensity
extends up to about 4.5 kpc.
The lower limit of about 1 kpc is less certain, due 
to incompleteness issues at the low-latitude side. The metallicity-distribution difference (bottom panel)
indicates the overdensity
has metallicities between -2.0 and -1.0 dex.  Its shape differs from that of the
{\it ABOVE} and {\it BELOW} distributions, having a pronounced wing at low metallicities.
A KS test performed between the difference distribution and the {\it BELOW} distribution
gives just a $16\%$ chance that they are drawn from the same population.

T15 search for overdensities in their entire survey
by RR Lyrae densities to predictions of a halo density model.
Their second most-significant overdensity is Hydra 1 (Hya 1; see Fig. 14 in T15) located in an area similar to ours,
but more extended in longitude. The longitude extent of our overdensity is limited by
the SPM4 declination limit reflected at $l\sim 0^{\circ}$, and by the kinematical analysis at the 
negative longitude end (see Section 3.3). Also, our overdensity is on average more distant 
than Hya 1: $\sim 7.8$ kpc compared to 5.1 kpc,
but its depth is similar: $\sim 4$ kpc. We believe that Hya 1 coincides with what we have detected here.
We note that our detection and
characterization is by direct comparison, above and below the plane,
while also including kinematics and metallicities.
   
\subsection{Kinematics from Transverse Velocities}

We calculate transverse velocities along longitude and latitude
as $V_{l,b} = k \times d \times \mu_{l,b}$  where $k=4.74$ is a constant to account 
for the appropriate units. Here, we use the extended samples, i.e., with the low-latitude limit of $|b| = 15^{\circ}$.
We also discard objects with $|V_{l,b}| > 700$ km/s. 
Besides comparing observed samples, we also use the Besancon galactic model (Robin et al. 2003)
to explore the expectations for a ``default'' Milky Way, along a grid of pointings as indicated in Fig. 1.
The Besancon simulations were run in the magnitude range
$10 < V < 18$, and assumed a fixed proper-motion error of 2 mas/yr. 
The simulated stars were then trimmed 
with the same distance and velocity cuts as applied to our observed samples.
For each pointing we calculate the average $V_l$ and $V_b$ for the entire sample, which we label ``all'',
and for a subsample with metallicities [Fe/H]$ \le -1.28$, which we label ``halo''. The ``all'' sample is
dominated by thin-disk kinematics, while the ``halo'' sample is dominated by halo kinematics with 
a small contribution from the thick disk.
This latter sample is chosen to be representative of our RR Lyrae stars. In the Besancon model,
the halo has 0 km/s rotation velocity, while the thick disk has 176 km/s. Thus our two Besancon samples 
act to guide us in $V_l$ space between a disk-like prograde population and a practically non-rotating, or mildly
prograde rotating population.

In Figure 3 we present the run of $V_l$ (left) and $V_b$ (right) as a function of longitude.
Besancon predictions are shown with colored triangles, while observations are shown with black symbols. 
Filled, large circles show velocity averages in longitude bins that contain 150 stars (except for the bin near $l=0^{\circ}$ 
which includes the remaining stars). The gray line shows a moving mean of the velocities. 
In the {\it BELOW} sample, Besancon predictions for the halo population agree reasonably well with the observations. 
However, in the {\it ABOVE} sample, both velocity components disagree with model predictions.
$V_b$  is offset by $\sim 70$ km/s from the prediction of nearly 0 km/s, 
across four longitude bins, while $V_l$ lies between the predictions for the thin disk and halo 
in the same longitude range. Thus, the overdensity has a net vertical motion away from the plane, 
and appears to be in a prograde motion, yet lagging the thin disk. 

\subsection{Checking Proper-Motion Systematics}

The overdensity's mean velocity
of $V_b \sim 70$ km/s seen in Fig. 3 (top, right) corresponds to about 2 mas/yr.
The {\it ABOVE} sample includes measurements from a total of 
23 SPM fields. It is unlikely that systematic errors in proper motions along RA and Dec would
produce a constant $\mu_b \sim 2$ mas/yr over some $25^{\circ}$ in the sky.

As a check, we make a comparison of the SPM4 to SPM2 --- a previous SPM version --- in 
the areas of interest. 
The SPM2 catalog contains only a fraction of the objects in the SPM4, but the proper motions
should be considered more reliable, as explained in G11.
In particular, in the SPM2 field-to-field systematics are better controlled,
due to the visual confirmation of all galaxies used as proper-motion reference.
Thus, even though the catalogs are not independent, the SPM2 can
still serve as a check on the SPM4.
The comparison shows that systematic errors of 1 mas/yr are the norm.
While regions with systematics up to 2 mas/yr do exist over short 
longitude spans, such are not seen in the $l=-25^{\circ}-0^{\circ}$ range of our samples.

Another check shown in Figure 4 is the run of $\mu_b$ as a function of $|Z|$ 
for the {\it ABOVE} sample. The dark line is a moving median of the 
data points. The average $\mu_b$ is about 2 mas/yr for vertical distances up to 4 kpc,
it then decreases to zero for $Z > 4.5$ kpc. 
This is the same $|Z|$ value at which the overdensity is no longer seen in the 
spatial distributions (Fig. 2, middle).
The above evidence argues against proper-motion 
systematics as an explanation for the intriguing kinematical feature seen in Fig. 3. 
Rather, we believe it is indicative of the kinematics
of this overdensity.

\section{Origin of the Overdensity}

\subsection{Orbital Constraints}

Knowing five of the six phase-space coordinates,
we now attempt to constrain the orbit of the overdensity {\it en masse}
by exploring plausible orbits for a ``test particle'' having the same mean properties.
First, we strive to isolate the sample of overdensity stars by trimming in parameter space 
where the overdensity is observed:
$l=-25^{\circ}-0^{\circ}$, $d=6-10$ kpc, $Z=1.5-4$ kpc, 
[Fe/H] = -2.0 to -1.0, and with proper-motion errors in both coordinates less 
than 5 mas/yr.
This yields a sample of
226 RR Lyrae from which we derive a mean position, distance, and proper motion -- the proper motion
being $\mu_l = -3.5$ mas/yr, $\mu_b = 1.9$ mas/yr.
A count of Catalina RR Lyrae stars in a symmetrically trimmed {\it BELOW} sample yields 119 stars, which
implies a rather large field contamination of roughly 53\% in our ``isolated'' overdensity sample.
Thus, a correction is made to the mean proper motion, assuming the field has 
$\mu_l = -4.5$ mas/yr, based on the similarly trimmed {\it BELOW} sample, and
$\mu_b = 0$ (reflex solar motion along $b$ is small at these distances).
After correcting for contamination, the overdensity-representing test particle is given a motion of
$\mu_l = -2.3$ mas/yr, $\mu_b = 4.1$ mas/yr.
It is located at $(X,Y,Z) = (1.0,-2.0,2.8)$ kpc,
where the Sun is at (8.0,0,0) kpc.
We adopt an uncertainty in the $\it mean$ proper motion of 1 mas/yr to 
account for possible systematics, which are described in Section 3.4.
Finally, we explore a range of values for the unknown heliocentric radial velocity;
from -200 km/s to +250 km/s, in steps of 50 km/s.

A set of 100 such orbit integrations were generated at each radial-velocity value using
a three-component analytical model of the Galaxy (Dinescu et al. 1999).
The results show that the test particle's orbit generally
exceeded the observed limits of the overdensity, 
namely $Z < 4$ kpc and $l > -25^{\circ}$, as seen in Figures 2 and 3.
Overall, only $8\%$ of the test orbits were contained within the observed spatial-distribution limits,
and none of these were from sets at the extremes in radial velocity, -200 and +250 km/s.

At the position of the overdensity -- 
$l \sim -15^{\circ}$, $b \sim +15^{\circ}$, and heliocentric distance $\sim8$ kpc --
$\mu_l$ is primarily aligned with velocity along Galactocentric radius,
line-of-sight velocity corresponds to rotational motion in the Galactic plane, and
$\mu_b$ correlates to vertical motion out of the plane.
With this geometry in mind, it becomes evident that the mean motion of the overdensity sample
{\it before} correction for possible field contamination was actually more likely to yield orbits
consistent with its observed spatial extent.
For instance, assigning the test particle the raw mean motion of the sample of 226 stars
and 
repeating the orbit integrations, yields
orbits that fall within the spatial limits $43\%$ of the time.
Low heliocentric-radial-velocity runs are still preferred, with $98\%$ of the orbits from sets
having $-150 < RV < +150$ km/s being consistent with the spatial distribution's limits.
Thus, either contamination is much lower than estimated, or proper-motion systematics have
conspired to shift the sample's mean motion by roughly 2-sigma.
Furthermore, these allowed orbits have low orbital angular momentum, $L_z$, with
values ranging from -300 to 700 km/s~kpc.
Radial-velocity measures of the candidate stars would decisively establish the 
prograde versus retrograde nature of the orbits.

\subsection{$\omega$ Centauri and Sagittarius as Progenitors}

The moderate vertical excursion of the stars in the overdensity (to $\sim 4$ kpc), 
and the small-size orbital angular momentum, allowing for both pro and
retrograde orbits, suggest that the overdensity may be debris from
$\omega$ Cen's parent satellite as envisioned by T04, 
Bekki \& Freeman (2003). While $\omega$ Cen is on a retrograde orbit (Dinescu et al. 1999),
N-body disruption models allow for 
a fraction of the debris to be on prograde orbits, especially closer to the center of the Galaxy.
Here, we explore the kinematics of the H4 model from 
T04, by ``observing'' an above-the-plane sample with 
the same spatial restrictions as in Section 3.3 ($l=-40^{\circ}-0^{\circ}$, $b=15^{\circ}-30^{\circ}$, and $1 \le d \le 15$ kpc).
In Table 1 we list the number of particles found within the spatial restrictions, the average $V_l$ and $V_b$.

We also explore the Law \& Majewski (2010) model for the disruption of 
the Sagittarius dwarf galaxy, applying the same restrictions as above; the respective quantities are
listed in the second line of the Table.

On the last line of Table 1 we present the same quantities, but for our best-trimmed sample of 
candidates in the overdensity as described in Section 4.1, to represent the kinematics of the overdensity.
Based on these specific models, the observations do not
confirm either an $\omega$ Cen or Sagittarius' origin.

Still, an $\omega$ Cen origin for the overdensity, in particular, cannot be ruled out. 
It is not a straightforward exercise to model the complex dynamical interaction of the central, 
non-axis symmetric parts of the Milky Way with the parent satellite of $\omega$ Cen.
Disruption models vary substantially
in the initial mass of the satellite, the launching location, 
and the time spent in disruption and phase-mixing.
For instance, the H4 (T04) $\omega$-Cen model predicts debris from the parent satellite 
above and below the plane, but with an excess below the plane. 
Presumably, had the launching location been 
flipped with respect to the plane, a spatial excess in the opposite sense would result.
Model predictions indicate a net positive heliocentric radial velocity for $\omega$ Cen debris
($\sim 50$ km/s),
different from the radial velocities of halo stars ($\sim -20$ km/s). Thus,
radial-velocity measures of the candidate overdensity stars and refined models will help to 
clarify the situation.  We note that velocity dispersion predicted by the H4 model
is large ($\sim 80-100$ km/s); thus, one should not expect to find a ``cold'' kinematical feature, as
is the case in more recent accretion events.


\begin{table}[tbh]
\caption{Kinematical Properties}
\begin{tabular}{llll}
\hline
\multicolumn{1}{c}{Sample} & \multicolumn{1}{c}{N} & \multicolumn{1}{c}{$<V_l>$} & \multicolumn{1}{c}{$<V_b>$} \\
 & & \multicolumn{1}{c}{(km/s)} & \multicolumn{1}{c}{(km/s)} \\
\hline
\multicolumn{1}{l}{$\omega$ Cen model} & \multicolumn{1}{c}{463} & \multicolumn{1}{r}{$-228\pm4$} & \multicolumn{1}{r}{$-14\pm5$} \\
\multicolumn{1}{l}{Sgr model} & \multicolumn{1}{c}{7} & \multicolumn{1}{r}{$-292\pm56$} & \multicolumn{1}{r}{$192\pm34$} \\
\multicolumn{1}{l}{RRab candidates} & \multicolumn{1}{c}{226} & \multicolumn{1}{r}{$-120\pm13$} & \multicolumn{1}{r}{$71\pm13$} \\
\hline
\end{tabular}
\end{table}


Finally, we point out
recent studies proposing resonant trapping in the disk and halo as the cause of some moving groups
and overdensities. Moreno et al. (2015) suggest that this trapping is induced by the
Galactic bar, while Molloy et al. (2015) suggest that bar-driven kinematic substructure can be found
near the bar as well as in the outer parts of the disk.
It is beyond the scope of this paper to examine such scenarios, but these might also offer
a viable origin explanation.  Although, the observed asymmetry, above-vs-below 
the plane, might be difficult to reconcile with effects from what is typically assumed to be a 
symmetric bar.

We thank Marcio Catelan for speedily providing us the SSS catalog of RR Lyrae stars while 
their publication was in press.
VIK acknowledges support from the grant N 231.01-214/013B to the Southern Federal University, Russia.

\begin{figure}
\includegraphics[scale=0.7,angle=-90]{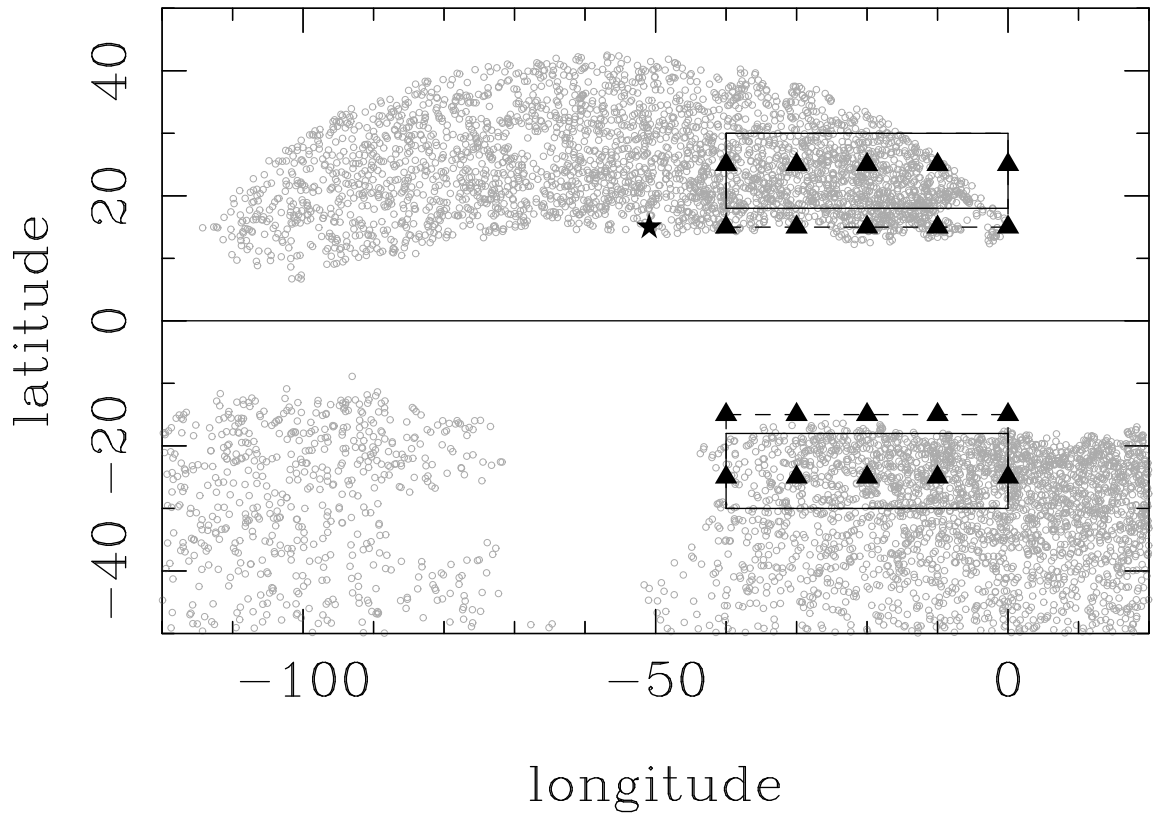}
\caption{Galactic-coordinate distribution of the SSS and SPM4 matched objects.
The rectangular boxes
show the areas of our two analyzed samples.
The ``primary'' samples' latitude range is $18^{\circ} \le |b| \le 30^{\circ}$ (continuous lines). 
The ``extended'' samples' latitude range is $15^{\circ} \le |b| \le 30^{\circ}$.
The Besancon samples' locations are shown with triangles, $\omega$ Cen's with a star.
}
\end{figure}

\begin{figure}
\includegraphics[scale=0.8,]{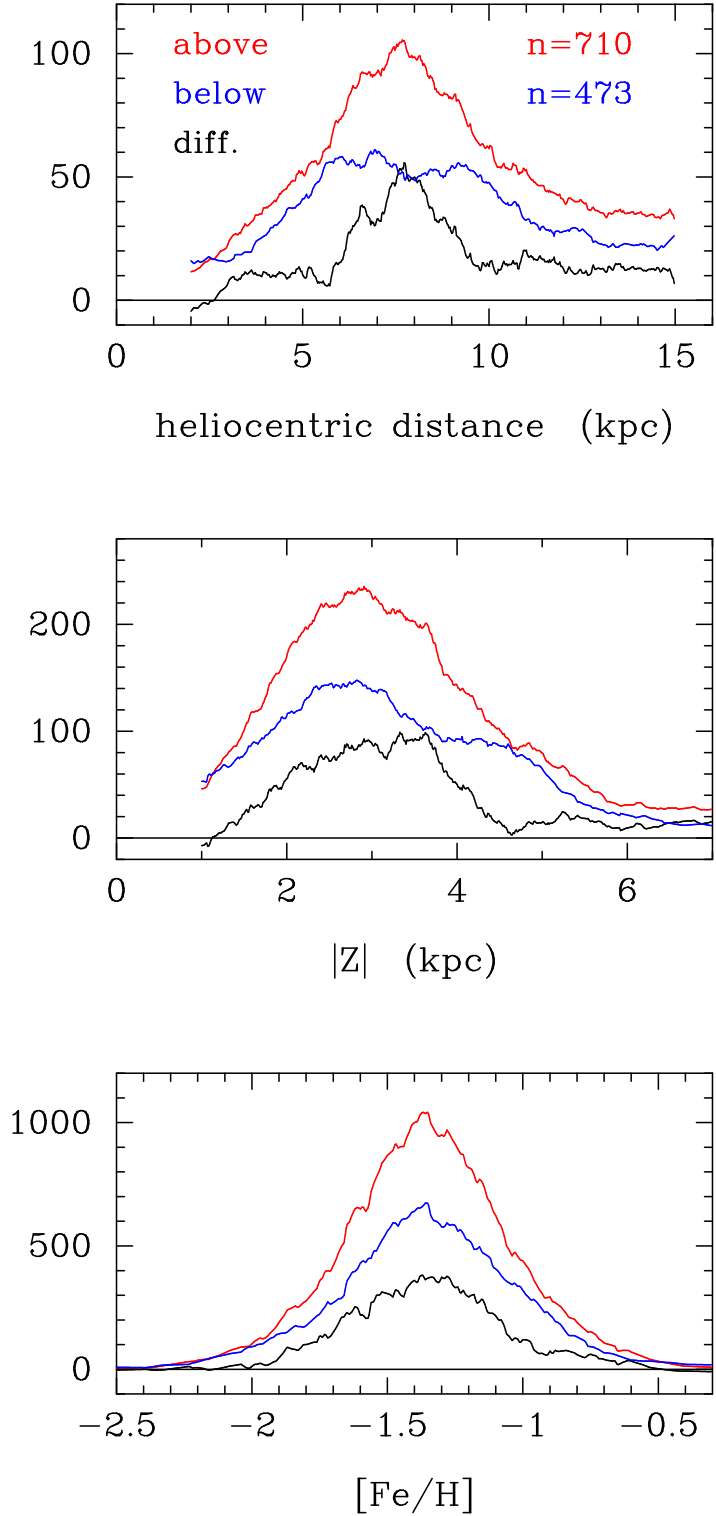}
\caption{Distributions in heliocentric distance (top), distance from the plane (middle),
and metallicity (bottom),
for the {\it ABOVE} and {\it BELOW} samples, and also the differences in the distributions.
Vertical scaling is number of stars per unit quantity along the abscissa.
}
\end{figure}

\begin{figure}
\includegraphics[scale=0.65,angle=-90]{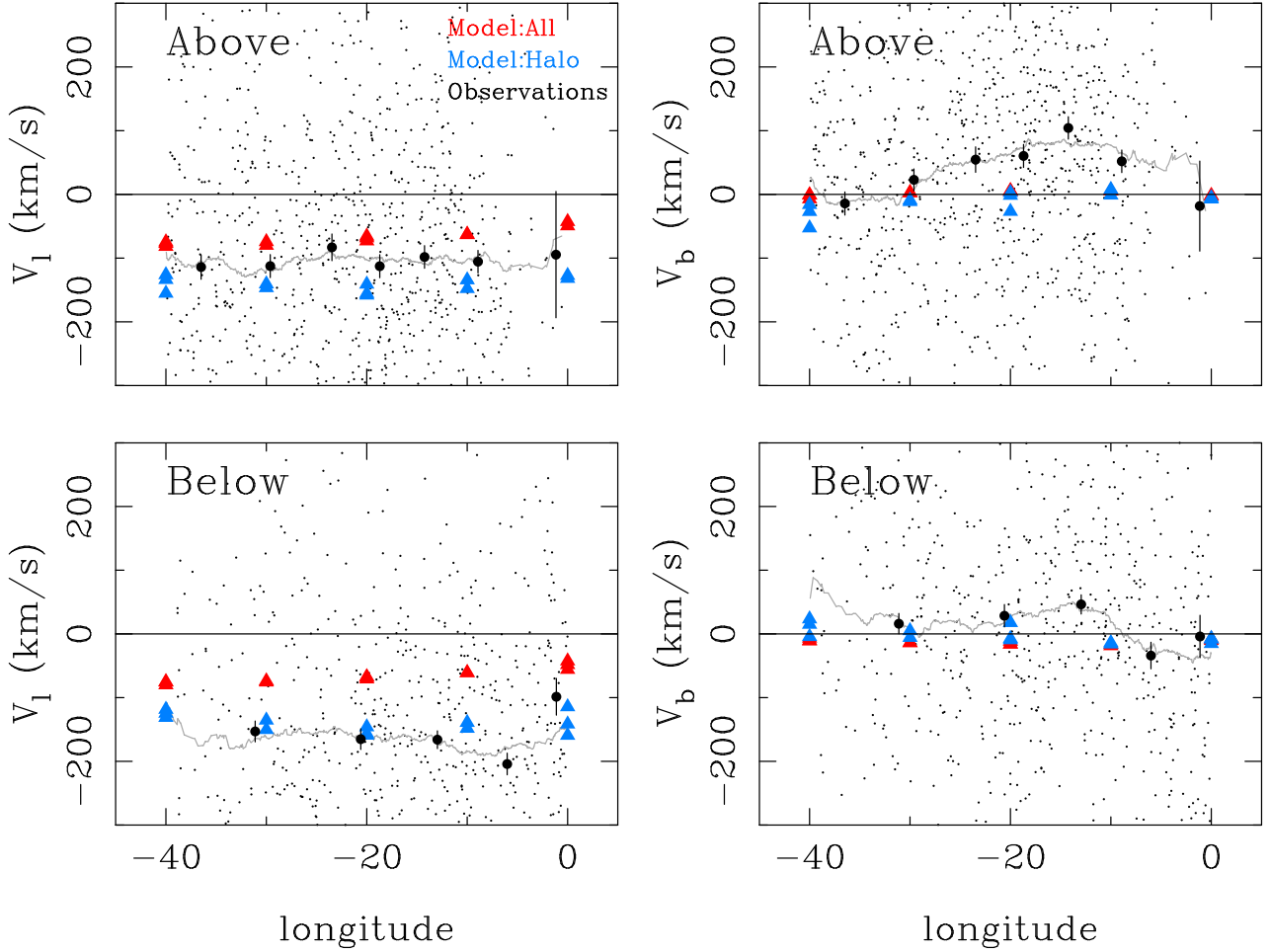}
\caption{Tangential velocities as a function of longitude for the {\it ABOVE} (top panels) and
{\it BELOW} (bottom panels) samples. 
Small dots are individual RRab stars. The gray lines represent moving means as a function of $l$.
Filled, black symbols show averages for bins of 150 RRab stars, except for the last bin near $l=0^{\circ}$, 
which includes the remaining stars in the sample. Triangles show predictions of the Besancon model
for the ``all'' sample (red) - representative of the thin disk, and for the ``halo'' sample representative
of the halo (blue) (see text).
}
\end{figure}

\begin{figure}
\includegraphics[scale=0.6,angle=-90]{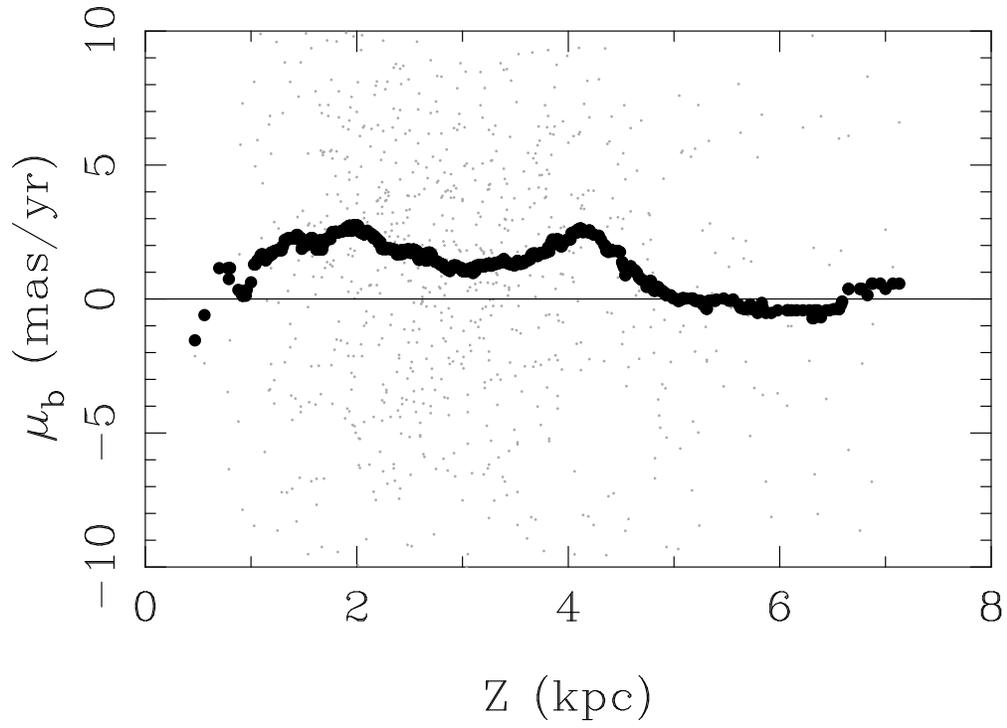}
\caption{Proper motion, $\mu_b$, as a function of $|Z|$
for the {\it ABOVE} sample $15^{\circ} \le b \le 30^{\circ}$. Small gray dots 
represent the RRab stars, while the black circles show a moving median.
}
\end{figure}

\end{document}